\documentclass[a4paper,11pt]{article}
\usepackage{pos}
\usepackage{natbib}
\usepackage{graphicx}
\usepackage{wrapfig}
\title{Improved Fermion Hamiltonians for Quantum Simulation}

\author[a,b]{Erik Gustafson}
\author[c]{Ruth Van de Water}

\affiliation[a]{University Space Research Association,\\
  Research Institute for Advanced Computer Science (RIACS), Mountain View CA}
\affiliation[b]{NASA Ames Research Center, \\ Quantum Artificial Intelligence Laboratory (QuAIL), Mountain View CA}
\affiliation[c]{Fermilab, Particle Theory Division\\
Kirk and Pine street, Batavia Il, USA}

\emailAdd{egustafson@usra.edu}
\emailAdd{ruthv@fnal.gov}

\abstract{We developed a Hamiltonian inspired by ASQTAD and highly improved staggered quark (HISQ) actions and show how these Hamiltonians can be used for quantum simulations. Gate costs for the time evolution of these improved Hamiltonians are provided as well as a demonstration of the reduction of lattice spacing errors using the 1+1d lattice Schwinger model.}

\FullConference{The 40th International Symposium on Lattice Field Theory (Lattice 2023)\\
July 31st - August 4th, 2023\\
Fermi National Accelerator Laboratory\\}

\begin{document}
\maketitle

\section{Introduction}
Improved actions were important for the accurate calculation of static quantities in Euclidean lattice gauge theories \cite{MILC:2009mpl}. 
These improvements allowed for calculations to be performed at larger lattice spacings. 
These developments typically followed the Symanzik improvement program which systematically adds terms to the Hamiltonian to cancel lattice errors at various orders. 
Two important actions that were developed for the simulation of euclidean lattice gauge theories including dynamical quarks were the ASQTAD and highly improved staggered quark (HISQ) Hamiltonians  \cite{PhysRevD.75.054502,Lepage:1997id,Lepage:1998vj,Lagae:1998pe}.

While Symanzik improvement offers many benefits to lattice calculations it cannot offer solutions to Sign problems that occur at finite density and with real-time evolution. 
Quantum computing offers a way to circumvent these sign problems entirely by using a Hamiltonian formalism for computation.
While the Kogut-Susskind Hamiltonian has been primarily used for recent quantum simulations, see \cite{Catterall:2022wjq} and references therein, study of improved Hamiltonians for quantum simulations has been sparser \cite{Carena:2022kpg,Carlsson:2003rf,Carlsson:2001wp,Luo:1998dx,ciavarella2023quantum}.

In this work we develop a method for quantum simulation of improved lattice gauge theories in a representation agnostic manner for ASQTAD and HISQ Fermions. The Hamiltonians are developed in Sec. \ref{sec:ASQTAD} and \ref{sec:HISQ}. The gate costs in terms of the primitive operations for gauge matter evolution ($\mathfrak{U}_{G.M}$), link inversion ($\mathfrak{U}_{-1}$), link multiplication ($\mathfrak{U}_{\times}$), electric field evolution ($\mathfrak{U}_{\rm{phase}}$), link trace phasing ($\mathfrak{U}_{Tr}$), electric-magnetic field basis transformations ($\mathfrak{U}_{F}$), and link reunitarization ($\mathfrak{U}_{R}$) are provided in Sec. \ref{sec:gatecosts}. A demonstration of the improved Hamiltonian for the model spectrum is provided in Sec. \ref{sec:example}.
\section{ASQTAD}
\label{sec:ASQTAD}
Staggered fermions naively have lattice discretization errors of order $\mathcal{O}(a^2)$.
The first step toward Symanzik improvement of the Hamiltonian is identifying which operators can be introduced to reduce the $\mathcal{O}(a^2)$ effects.
The most straight forward term to incorporate is the ``Naik" term \cite{Naik:1986bn} which corrects for tree level errors in the derivative operator.
This involves adding a correction term to the derivative operator,
\begin{equation}
    \label{eq:naikterm}
    \Delta_{m}[U] \rightarrow \Delta_m[U] - \frac{1}{6}\Delta^3_m[u].
\end{equation}
The second source of $\mathcal{O}(a^2)$ errors comes from ``taste" exchange interactions \cite{Toussaint:1998sa,Orginos:1998ue, Lepage:1997id,Lagae:1998pe,Lepage:1998vj} that couple low momentum gauge bosons to high momentum gauge bosons near the corners of the Brillouin zone. 
These taste exchange interactions can be removed by adding in four fermion contact terms \cite{Lepage:1997id,Lepage:1998vj,Lagae:1998pe} or by smearing the gauge links \cite{Hasenfratz:2001hp}.
Adding additional four quark operators to the Hamiltonian is non-trivial and invovles designing many new primitive gates for Trotter operations.
These operators will also change the time dependent lattice spacing discretization effects from Trotterization.
Therefore, a quantum smearing algorithm is desirable to circumvent this problem, such as in Ref. \cite{Gustafson:2022hjf}, is desirable.

In order to write down the ASQTAD Hamiltonian it is worth following a parallel derivation of the ASQTAD action from Ref. \cite{PhysRevD.75.054502}.
In the action formalism, one first averages the target link $U$ appearing in the single link hopping term with the neighboring staples containing 3, 5, and 7 links such that the $\mathcal{O}(a^2)$ effects are suppressed. 
In the Hamiltonian picture this corresponds to the link transformation
\begin{equation}
    \label{eq:quantumsmearing}
    \hat{U}_m(\vec{r}) \rightarrow \Bigg(\mathcal{F}^{(1)} + \mathcal{F}^{(2)}\Bigg)\hat{U}_m(\vec{r}) = \mathcal{F}^{\rm{ASQTAD}}[\hat{U}_m(\vec{r})].
\end{equation}
The operators are defined as 
\begin{equation}
    \label{eq:asqtadF1}
    \mathcal{F}^{(1)} = \Bigg(\prod_{n\neq m}(1 + \frac{a^2\Delta^{(2)}_{n}}{4}\Bigg)|_{\rm{symm.}}~\text{and}~
    \mathcal{F}^{(2)}= -\frac{a^2}{4} \sum_{n\neq m}(\Delta_n)^2,
\end{equation}
where the symmetrization indicates the derivatives are taken across all possible permutations.
The operators $\Delta_n$ and $\Delta^{(2)}_n$ correspond first and second order coviatian derivatives and are defind as follows:
\begin{equation}
    \Delta_n \hat{U}_m(\vec{r}) = \frac{1}{a}\big(\hat{U}_n(\vec{r}) \hat{U}_{m}(\vec{r} + \hat{n}) \hat{U}^{\dagger}(\vec{r} + \hat{m}) - \hat{U}^{\dagger}_n(\vec{r} - \hat{n}) \hat{U}_{m}(\vec{r} - \hat{n}) \hat{U}(\vec{r} + \hat{m} - \hat{n})\big)
\end{equation}
and
\begin{equation}
    \Delta_n^{(2)} \hat{U}_m(\vec{r}) = \frac{1}{a}( \hat{U}_n(\vec{r}) \hat{U}_{m}(\vec{r} + \hat{n}) \hat{U}^{\dagger}(\vec{r} + \hat{m}) - 2\hat{U}_m(\vec{r}) + \hat{U}^{\dagger}_n(\vec{r} - \hat{n}) \hat{U}_{m}(\vec{r} - \hat{n}) \hat{U}(\vec{r} + \hat{m} - \hat{n}).
\end{equation}
An explicit evaluation of the smearing procedure for gauge links in Eq. (\ref{eq:quantumsmearing}) is omitted but can be derived with a modicum of algebra.
An important aspect for quantum computation is tha tthe link \emph{must} be reunitarized denoted by the functional $\mathcal{R}[\hat{U}]$.
Because a linear combination of states is not equivalent to a state corresponding to the average of the linear combination of elements.
This involves projecting the linear combination of links back onto a group element such as was developed in \cite{Gustafson:2022hjf}.
We are now free to write down the AQSTAD portion of the Hamiltonian corresponding to the matter terms:
\begin{equation}
        \label{eq:astadham}
        \begin{split}
        H_{\text{kinetic}}^{ASQTAD} = &\frac{1}{2 a} \sum_{\vec{r}}\sum_{n} \Big\lbrace \chi_m(\vec{r}) \phi^{\dagger}(\vec{r})\bigg(\mathcal{R}[\mathcal{F}^{ASQTAD}[U_m(\vec{r})]] \phi(\vec{r}+\hat{m})\\
        & -\frac{1}{24}\hat{U}_m(\vec{r})\hat{U}_m(\vec{r} + \hat{m})\hat{U}_m(\vec{r} + 2\hat{m}) \phi(\vec{r} + 3 \hat{m})\bigg) + h.c.\Big\rbrace.
        \end{split}
\end{equation}
\section{HISQ}
\label{sec:HISQ}
The ASQTAD action and Hamiltonian address the tree level $\mathcal{O}(a^2)$ effects but do not address the loop errors $\mathcal{O}(\alpha_s a^2)$ errors which can be non-negligible.
In the action formalism this was addressed by implementing a second level of ASQTAD smearing on the one link terms and a first round of ASQTAD smearing on the ``Naik" term. This resulted in the HISQ action\cite{PhysRevD.75.054502}.
Following the same method as the ASQTAD derivation, the HISQ Hamiltonians Kinetic term will be
\begin{equation}
    \label{eq:hamhisq}
            \begin{split}
            H^{HISQ}_{kinetic} = &\frac{1}{2a} \sum_{\vec{r},m} \bigg\lbrace \xi_{m}(\vec{r}) \phi^{\dagger}(\vec{r}) \Big( \hat{W}_m(\vec{r}) \phi(\vec{r} + \hat{m}) \\ & -\frac{1}{24} \hat{X}_m(\vec{r})\hat{X}_m(\vec{r}+\hat{m})\hat{X}_m(\vec{r}+2\hat{m})\phi(\vec{r}+3\hat{m})\Big) + h.c.\bigg\rbrace
            \end{split}
\end{equation}
where
\begin{equation}
    \hat{W}_m(\vec{r}) = \mathcal{R}[\mathcal{F}^{\rm{HISQ}}[\hat{U}_m(\vec{r})]] \text{ and } \hat{X}_m(\vec{r}) = \mathcal{R}[\mathcal{F}^{(1)}[\hat{U}_m(\vec{r})]],
\end{equation}
with
\begin{equation}
\mathcal{F}^{HISQ}[\hat{U}_m(\vec{r})] = \Big(\mathcal{F}^{(1)}[\mathcal{R}[\mathcal{F}^{(1)}[\hat{U}_m(\vec{r})]]] - 2 \mathcal{F}^{(2)}[\mathcal{R}[\mathcal{F}^{(1)}[\hat{U}_m(\vec{r})]]]\Big).
\end{equation}
\section{Implementation and Gate Costs}
\label{sec:gatecosts}
We need to express how these operations work as an algorithm for Trotterization and time evolution. We first discuss the ASQTAD Hamiltonian and follow with a summary for the HISQ Hamiltonian.

For the ASQTAD Hamiltonian, we first multiply the 3 links together that correspond to the ``Naik" term. Then the primitive gauge kinetic hopping evolution operator is applied on this ``Naik" term including appropriate antisymmetrization rules for the fermions. The next step is to calculate the staples $S^{(3)}_{n,m}(\vec{r})$, $S^{(5)}_{l,n,m}(\vec{r})$, and $S^{(5)}_{n, n, m}(\vec{r})$ onto scratch link registers.
One then follows the proceedure in Ref. \cite{Gustafson:2022hjf} to calculate $\mathcal{R}[\mathcal{F}^{\rm{ASQTAD}}[\hat{U}_{m}(\vec{r})]]$ using the previously calculated staples.
Next we apply the primitive hopping term using the smeared link.
Finally all the reunitarized gauge link and staples are uncomputed.

The HISQ evolution term follows similarly. First we calculate the staples $S^{(3)}_{n, m}$ and $S^{(5)}_{l, n, m}$ onto scratch registers and then compute $\mathcal{R}[\mathcal{F}^{(1)}[\hat{U}_m]]$ onto a new lattice worth of qubits denoted $\Lambda_X$, these are the $\hat{X}_m(\vec{r})$ onto an additional set of scratch registers we compute $\hat{S}^{(3)}_{n, m}$,$S^{(5)}_{l,n,m}(\vec{r})$, and $S^{(5)}_{n, n, m}(\vec{r})$ from $\Lambda_X$. 
One now computes $\mathcal{R}[\mathcal{F}^{(1)}[\hat{X}_{m}(\vec{r})] - 2 \mathcal{F}^{(2)}[\hat{X}_{m}(\vec{r})]]$ onto a new set of scratch registers denoted $\Lambda_W$.
The one link kinetic operators are implemented using the $\Lambda_W$ links, while the ``Naik" term operation is applied using the $\Lambda_X$ links. 
Finally one computes $\Lambda_W$, $\Lambda_X$ and its corresponding staples, and the staples from the original lattice. 
The total gate costs in terms of primitive group operations are provided in Tab. \ref{tab:gatecosts}
\begin{table}[!ht]
\caption{Number of primitive group operations per link register per Trotter step for the Fermionic terms. $K.S.$ is the Kogut-Susskind costs, $NR$ indicates the cost of not reunitarizing the gauge links, and $RE$ indicates the cost of reunitarizing the gauge links.}
\label{tab:gatecosts}
\begin{tabular}{cccccc}
\hline\hline
Gate & Naive K.S. & $O(a^2)$ gauge & ASQTAD NR & Asqtad RE & HISQ\\
\hline
$\mathfrak U_{G.M.}$ & 1 & 0 & $14(d - 1) - 11$ & 2 & 2\\
$\mathfrak U_{-1}$ & $3(d-1)$ & $2 + 8(d - 1)$ & $52(d - 1) - 48$ & $52(d - 1) - 48$ & $104(d - 1) - 96$\\
$\mathfrak U_{\times}$ & $6(d-1)$& $4+20(d-1)$ & $132d - 256$ & $132d - 256$ & $264d - 512$\\
$\mathfrak U_{\text{phase}}$ & 1& 1 & 0 & 0& 0\\
$\mathfrak U_{Tr}$ & $\frac{d - 1}{2}$& $d-1$ & 0 & 0& 0\\
$\mathfrak U_{F}$ & 2 & 2 & 0 & 0& 0 \\
$\mathfrak U_{U}$ & 0 & 0 & 0 & 2 & 4\\
\hline\hline
\end{tabular}
\end{table}
\section{Example case: the Schwinger model}
\label{sec:example}
While the greatest effects from the ASQTAD and HISQ Hamiltonians will be seen in 2 and 3 spacial dimensions since this is where taste splitting will be apparent, certain effects of the improved Hamiltonian will be visible in 1 spacial dimension. For this reason we pick the Schwinger model as a test bed and examine the low lying spectrum. 
For brevity the Hamiltonian is written after the gauge fields have been integrated out, and a Jordan-Wigner transformation has been performed to write the Hamiltonian in terms of spin operators.
The naive Kogut-Susskind, $H_{k.s.}$ and ASQTAD, $H_{imp.}$, Hamiltonians respectively are
\begin{equation}
    H_{k.s} =  H_{m.g.} + H_{E} + H_{m}~\text{and}~H_{imp.} =  H_{m.g.} + H_{Naik} + \frac{5}{6} H_E + \frac{1}{6} H_E^{(2)} + H_m
\end{equation}
where
\begin{equation*}
\begin{split}
H_{m.g.} = &\frac{1}{4 a} \sum_{v} (X_{v}X_{v + 1} + Y_{v}Y_{v+1}),~H_E =  \frac{g^2}{8} \sum_{n=0}^{N-2}\sum_{v,w=1}^{n}\bigg(Z_v Z_w + (-1)^{v + 1}Z_w + (-1)^{w + 1} Z_v\bigg),\\  H_m =& \frac{m_{lat}}{2} \sum_{v} (-1)^{v}Z_v,~H_{E}^{(2)} = \frac{1}{4} \sum_{n=0}^{N-3}\sum_{v=0}^{n}\sum_{w=0}^{n+1} Z_{v}Z_{w} + (-1)^{v + 1}Z_{w} + (-1)^{w + 1} Z + (-1)^{v + w},\\
\end{split}\end{equation*}
\begin{equation*}
    \begin{split}
H_{Naik} =  &\frac{1}{48} (X_0 X_1 + Y_0 Y_1 + X_{N-2} X_{N-1} + Y_{N-2}Y_{N-1}) +\frac{3}{96}\sum_{v=1}^{N-3}(X_n X_{n+1} + Y_{n} Y_{n+1})\\
    & + \frac{-1}{96} \sum_{v=0}^{N-4}( X_{n}Z_{n+1}Z_{n+2}X_{n+3} + Y_{n}Z_{n+1}Z_{n+2}Y_{n+3}),\\
\end{split}
\end{equation*}
$N$ is the number of lattice sites, $g^2$ is the gauge field coupling, $a$ the lattice spacing which is set to $1$, and $m_{lat}$ is the bare fermion mass. 

\begin{wrapfigure}{r}{4in}
\includegraphics[width=4in]{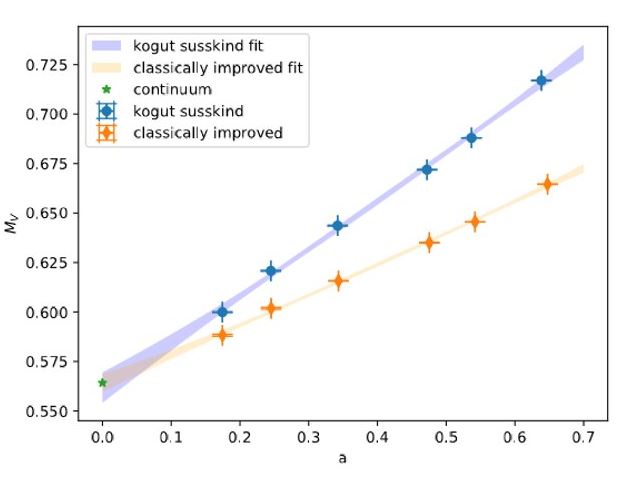}
\caption{Continuum Limit extrapolation for the Vector mass using the improved and Naive Kogut-Susskind Hamiltonian}
\label{fig:continuumcalc}
\end{wrapfigure}
We show the continuum limit extrapolation for the vector excitation of the Schwinger model in Fig. \ref{fig:continuumcalc}. 
A striking result is that the slope of the vector mass is shallower indicating substantial portions of the Tree-level errors have been removed however there still appear to be non trivial effects present.
In particular there could be effects related to the electric field in the gauge matter term, non-negligible effects from one loop corrections, and renormalization of the fermion mass \cite{dempsey2023discrete}.
Each of these effects are planned studies for further investigations. 
\section{Outlook}
In this work we have demonstrated a method to construct a ASQTAD and HISQ Hamiltonian suitable for the simulation of lattice gauge theories. These Hamiltonians in principle should reduce the effects of lattice discretization and in the future allow for calculations of dynamical quantities in quantum field theories with fewer quantum resources than the Kogut-Susskind Hamiltonian alone on fault-tolerant quantum computing hardware.
As improved quantum hardware becomes available, it will be important to test aspects of these Hamiltonians with toy models such as $\mathbb{Z}_n$ and nonabelian gauge theories in 2 and 3 spacial dimensions where the effects of taste splitting will become important. 

\begin{acknowledgments}
    We thank Benoit Assi, Tom Cohen, Adrien Florio, Florian Herren, Andreas Kronfeld, Rob Pisarski, and Norman Tubman for helpful comments. This work is supported by the Department of Energy through the National Quantum Information Science Research Centers, Superconducting Quantum Materials and Systems Center (SQMS) under the contract No. DE-AC02-07CH11359 (E.G.) and Fermilab QuantiSED program in the area of ``Intersections of QIS and Theoretical Particle Physics" (RV). E.G. was supported by the NASA Academic Mission Services, Contract No. NNA16BD14C.
\end{acknowledgments}

\bibliographystyle{JHEP}
\bibliography{apssamp.bib}

\providecommand{\noopsort}[1]{}\providecommand{\singleletter}[1]{#1}%

\providecommand{\href}[2]{#2}\begingroup\raggedright\begin{thebibliography}{10}

\bibitem{MILC:2009mpl}
{\scshape MILC} collaboration, \emph{{Nonperturbative QCD Simulations with 2+1
  Flavors of Improved Staggered Quarks}},
  \href{https://doi.org/10.1103/RevModPhys.82.1349}{\emph{Rev. Mod. Phys.}
  {\bfseries 82} (2010) 1349}
  [\href{https://arxiv.org/abs/0903.3598}{{\ttfamily 0903.3598}}].

\bibitem{PhysRevD.75.054502}
E.~Follana, Q.~Mason, C.~Davies, K.~Hornbostel, G.P.~Lepage, J.~Shigemitsu
  et~al., \emph{Highly improved staggered quarks on the lattice with
  applications to charm physics},
  \href{https://doi.org/10.1103/PhysRevD.75.054502}{\emph{Phys. Rev. D}
  {\bfseries 75} (2007) 054502}.

\bibitem{Lepage:1997id}
P.~Lepage, \emph{{Perturbative improvement for lattice QCD: An Update}},
  \href{https://doi.org/10.1016/S0920-5632(97)00489-1}{\emph{Nucl. Phys. B
  Proc. Suppl.} {\bfseries 60} (1998) 267}
  [\href{https://arxiv.org/abs/hep-lat/9707026}{{\ttfamily hep-lat/9707026}}].

\bibitem{Lepage:1998vj}
G.P.~Lepage, \emph{{Flavor symmetry restoration and Symanzik improvement for
  staggered quarks}},
  \href{https://doi.org/10.1103/PhysRevD.59.074502}{\emph{Phys. Rev. D}
  {\bfseries 59} (1999) 074502}
  [\href{https://arxiv.org/abs/hep-lat/9809157}{{\ttfamily hep-lat/9809157}}].

\bibitem{Lagae:1998pe}
J.F.~Lagae and D.K.~Sinclair, \emph{{Improved staggered quark actions with
  reduced flavor symmetry violations for lattice QCD}},
  \href{https://doi.org/10.1103/PhysRevD.59.014511}{\emph{Phys. Rev. D}
  {\bfseries 59} (1999) 014511}
  [\href{https://arxiv.org/abs/hep-lat/9806014}{{\ttfamily hep-lat/9806014}}].

\bibitem{Catterall:2022wjq}
S.~Catterall et~al., \emph{{Report of the Snowmass 2021 Theory Frontier Topical
  Group on Quantum Information Science}},  in \emph{{Snowmass 2021}}, 9, 2022
  [\href{https://arxiv.org/abs/2209.14839}{{\ttfamily 2209.14839}}].

\bibitem{Carena:2022kpg}
M.~Carena, H.~Lamm, Y.-Y.~Li and W.~Liu, \emph{{Improved Hamiltonians for
  Quantum Simulations of Gauge Theories}},
  \href{https://doi.org/10.1103/PhysRevLett.129.051601}{\emph{Phys. Rev. Lett.}
  {\bfseries 129} (2022) 051601}
  [\href{https://arxiv.org/abs/2203.02823}{{\ttfamily 2203.02823}}].

\bibitem{Carlsson:2003rf}
J.~Carlsson, \emph{{Improvement and analytic techniques in Hamiltonian lattice
  gauge theory}},  other thesis, 4, 2003,
  [\href{https://arxiv.org/abs/hep-lat/0309138}{{\ttfamily hep-lat/0309138}}].

\bibitem{Carlsson:2001wp}
J.~Carlsson and B.H.J.~McKellar, \emph{{Direct improvement of Hamiltonian
  lattice gauge theory}},
  \href{https://doi.org/10.1103/PhysRevD.64.094503}{\emph{Phys. Rev. D}
  {\bfseries 64} (2001) 094503}
  [\href{https://arxiv.org/abs/hep-lat/0105018}{{\ttfamily hep-lat/0105018}}].

\bibitem{Luo:1998dx}
X.-Q.~Luo, S.-H.~Guo, H.~Kroger and D.~Schutte, \emph{{Improved lattice gauge
  field Hamiltonian}},
  \href{https://doi.org/10.1103/PhysRevD.59.034503}{\emph{Phys. Rev. D}
  {\bfseries 59} (1999) 034503}
  [\href{https://arxiv.org/abs/hep-lat/9804029}{{\ttfamily hep-lat/9804029}}].

\bibitem{ciavarella2023quantum}
A.N.~Ciavarella, \emph{Quantum simulation of lattice qcd with improved
  hamiltonians},  2023.

\bibitem{Naik:1986bn}
S.~Naik, \emph{{On-shell Improved Lattice Action for {QCD} With Susskind
  Fermions and Asymptotic Freedom Scale}},
  \href{https://doi.org/10.1016/0550-3213(89)90394-5}{\emph{Nucl. Phys. B}
  {\bfseries 316} (1989) 238}.

\bibitem{Toussaint:1998sa}
{\scshape MILC} collaboration, \emph{{Tests of improved Kogut-Susskind fermion
  actions}}, \href{https://doi.org/10.1016/S0920-5632(99)85241-4}{\emph{Nucl.
  Phys. B Proc. Suppl.} {\bfseries 73} (1999) 909}
  [\href{https://arxiv.org/abs/hep-lat/9809148}{{\ttfamily hep-lat/9809148}}].

\bibitem{Orginos:1998ue}
{\scshape MILC} collaboration, \emph{{Testing improved actions for dynamical
  Kogut-Susskind quarks}},
  \href{https://doi.org/10.1103/PhysRevD.59.014501}{\emph{Phys. Rev. D}
  {\bfseries 59} (1999) 014501}
  [\href{https://arxiv.org/abs/hep-lat/9805009}{{\ttfamily hep-lat/9805009}}].

\bibitem{Hasenfratz:2001hp}
A.~Hasenfratz and F.~Knechtli, \emph{{Flavor symmetry and the static potential
  with hypercubic blocking}},
  \href{https://doi.org/10.1103/PhysRevD.64.034504}{\emph{Phys. Rev. D}
  {\bfseries 64} (2001) 034504}
  [\href{https://arxiv.org/abs/hep-lat/0103029}{{\ttfamily hep-lat/0103029}}].

\bibitem{Gustafson:2022hjf}
E.J.~Gustafson, \emph{{Stout Smearing on a Quantum Computer}}, {\emph{arxiv}
  (2022) } [\href{https://arxiv.org/abs/2211.05607}{{\ttfamily 2211.05607}}].

\bibitem{dempsey2023discrete}
R.~Dempsey, I.R.~Klebanov, S.S.~Pufu and B.~Zan, \emph{Discrete chiral symmetry
  and mass shift in lattice hamiltonian approach to schwinger model},  2023.

\end{thebibliography}\endgroup


\end{document}